# A Lunar Farside Low Radio Frequency Array for Dark Ages 21-cm Cosmology

**Authors:** Jack O. Burns (University of Colorado Boulder), Gregg Hallinan (California Institute of Technology), Tzu-Ching Chang (Jet Propulsion Laboratory/Caltech), Marin Anderson (JPL/Caltech), Judd Bowman (ASU), Richard Bradley (NRAO), Steven Furlanetto (UCLA), Alex Hegedus (U. Michigan), Justin Kasper (U. Michigan), Jonathon Kocz (Caltech), Joseph Lazio (JPL/Caltech), Jim Lux (JPL/Caltech), Robert MacDowall (NASA Goddard), Jordan Mirocha (McGill U.), Issa Nesnas (JPL/Caltech), Jonathan Pober (Brown U.), Ronald Polidan (Lunar Resources), David Rapetti (ARC/USRA/CU), Andres Romero-Wolf (JPL/Caltech), Anže Slosar (Brookhaven National Laboratory), Albert Stebbins (Fermilab), Lawrence Teitelbaum (JPL/Caltech), Martin White (UC Berkeley/Lawrence Berkeley National Laboratory)

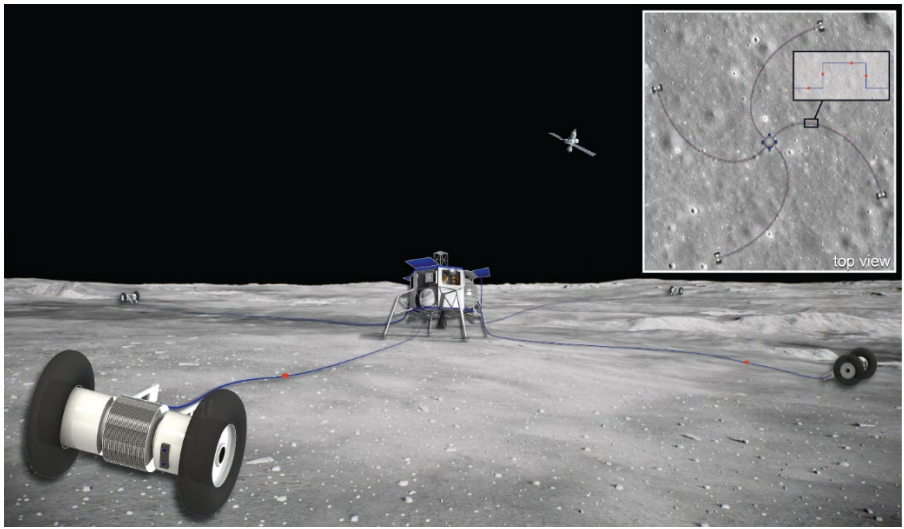

*A NASA Probe mission study produced a viable engineering design for an initial farside radio array that consists of 3 components: a commercial lander carrying the base station, four JPL Axel rovers to deploy antenna nodes, and a 128×2 (two orthogonal polarizations) node antenna array. The array will be deployed in a spiral pattern as shown in the upper right. Tethers connect the lander to the nodes, providing communications and power. NASA's Gateway may serve as a data relay to Earth.*

**ABSTRACT – Focus Area 1: lunar farside radio telescope to explore the early universe**

An array of low-frequency dipole antennas on the lunar farside surface will probe a unique, unexplored epoch in the early Universe called the Dark Ages. It begins at Recombination when neutral hydrogen atoms formed, first revealed by the cosmic microwave background. This epoch is free of stars and astrophysics, so it is ideal to investigate high energy particle processes including dark matter (e.g., warm dark matter, self-annihilation, nongravitational interactions with baryons), early Dark Energy, neutrinos, and cosmic strings. A NASA-funded study investigated the design of the instrument and the deployment strategy from a lander of 128 pairs of antenna dipoles across a 10 km×10 km area on the lunar surface. The antenna nodes are tethered to a base station/lander for central data processing, power, and data transmission to a relay satellite. The resultant architecture is broadly consistent with the requirements of a pathfinder Probe mission to begin these observations. The array, named FARSIDE, would provide the capability to image the entire sky in 1400 channels spanning frequencies from 100 kHz to 40 MHz, extending down two orders of magnitude below bands accessible to ground-based radio astronomy. The lunar farside can simultaneously provide isolation from terrestrial radio frequency interference, the Earth's auroral kilometric radiation, and plasma noise from the solar wind. It is thus the only location within the inner solar system from which sky noise limited observations can be carried out at sub-MHz frequencies. Through precision calibration via an orbiting beacon and exquisite foreground characterization, the farside array would measure the Dark Ages global 21-cm signal at redshifts $z \sim 35\text{-}200$. It will also be a pathfinder for a larger 21-cm power spectrum instrument by carefully measuring the foreground with high dynamic range.

## 1. Key Science Opportunity

In this RFI response, a first-generation radio interferometer consisting of an array of 128×2 dipole antennas deployed over a 10-km×10-km area on the lunar farside, named FARSIDE, is described as a new tool for investigating the Dark Ages epoch in the early Universe. FARSIDE enables precision measurements of the highly redshifted 21-cm signal from neutral hydrogen at frequencies ∼5–40 MHz, corresponding to redshifts z≈236-35, and provides a unique probe of an otherwise inaccessible epoch of the Universe. Such observations have enormous potential to enable new powerful tests of the standard cosmological model in the early Universe. They would provide constraints on any exotic physics operating at early times, including dark matter annihilation, nongravitational interaction with baryons, warm or fuzzy dark mater, early Dark Energy, sterile neutrinos, cosmic strings, etc. Additionally, FARSIDE would pave the way for a future generation of large radio arrays on the lunar farside, to access the ultimate number of observable modes in the Universe. It would be capable of exquisite measurements in fundamental physics and cosmology, including the spatial curvature of the Universe, neutrino masses and hierarchy, and the physics of inflation. These topics are well-aligned with DOE HEP and NASA APD drivers.

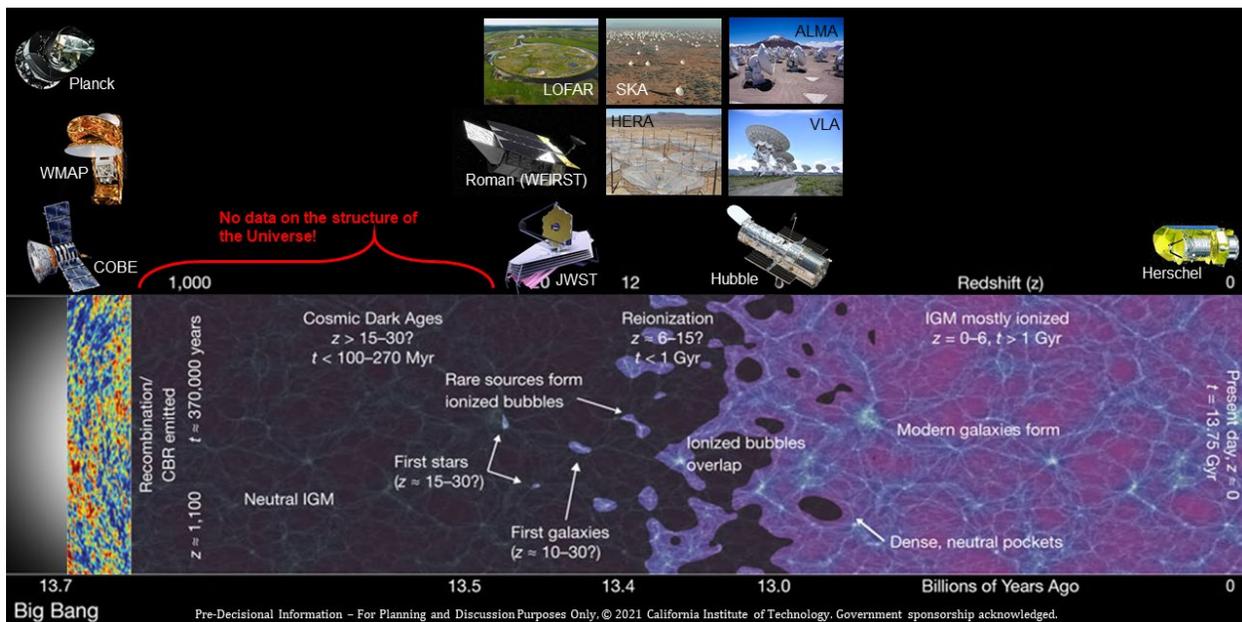

**Figure 1:** The pre-stellar (Dark Ages), first stars (Cosmic Dawn), and Reionization epochs of the Universe can be uniquely probed using the redshifted 21-cm signal. This history is accessible via the neutral hydrogen spin-flip background. Credit: JPL/Caltech (Bandyopadhyay et al. 2021).

Figure 1 places the Dark Ages, Cosmic Dawn (first stars), and Epoch of Reionization (EoR) into perspective. After the Big Bang, the Universe was hot, dense, and nearly homogeneous. As the Universe expanded, the material cooled, condensing after ∼400,000 years (z∼1100) into neutral atoms, freeing the Cosmic Microwave Background (CMB) photons. The baryonic content during this pre-stellar Dark Ages of the Universe consisted primarily of neutral hydrogen. About fifty million years later, gravity propelled the formation of the first luminous objects – stars, black holes, and galaxies – which ended the Dark Ages and commenced the Cosmic Dawn (e.g., Loeb and Furlanetto 2013). These first stars (Pop III, nearly metal-free) likely differed dramatically from stars we see nearby, as they formed in vastly different environments (Abel et al. 2002).

A farside array will employ two complementary approaches to observe the Dark Ages (z∼236−35, ∼4.4−80 Myrs after the Big Bang) using the highly redshifted 21-cm signal from neutral hydrogen, the most ubiquitous baryonic matter in the Universe. First, the sky-averaged Global signal or monopole can be observed by a spectrometer connected to individual dipole antennas operating in the total power mode. This is analogous to the CMB blackbody spectrum that was precisely measured by COBE. But, unlike the CMB, the spectral 21-cm line enables the redshift evolution of the hydrogen abundance and temperature to be measured (Figure 2). The observed brightness temperature (power over frequency bandwidth), $T_b(\nu)$, is a gauge of the evolution of the neutral hydrogen density, along with the radio background and gas temperature. In the Dark Ages before any luminous sources turn-on, these physical parameters depend only upon the cosmic adiabatic expansion of the Universe. Any deviation from the standard cosmological model requires novel additional physics. Therefore, a measurement of the hydrogen Global signal is not only a powerful test of the standard cosmological model in an epoch heretofore unobserved but also a flag of exotic physics in the early Universe (Furlanetto et al. 2019).

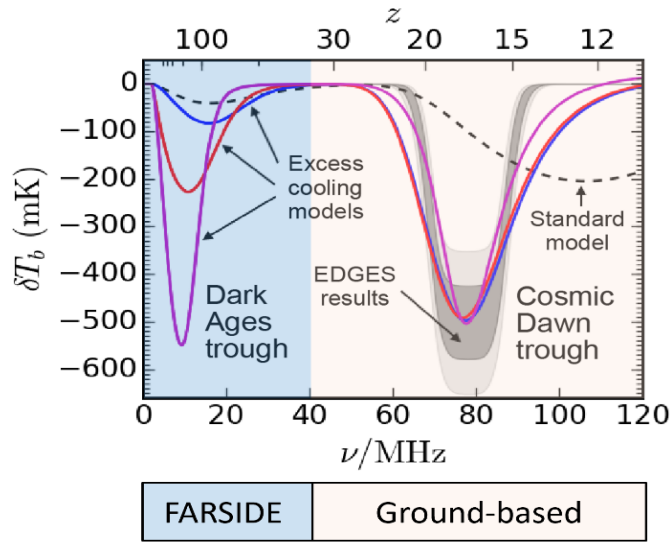

**Figure 2. The Dark Ages 21-cm absorption trough is a sensitive probe of cosmology.** The black dashed curve shows the brightness temperature (relative to the radio background) in a standard cosmological model with adiabatic cooling. The shape at z > 30 is independent of astrophysical sources. The grey contours show the 1 & 2-σ absorption bands inferred by EDGES (Bowman et al. 2018). The solid color curves are parametric models that invoke extra cooling to match the amplitude of the EDGES signal in the Cosmic Dawn frequency range but can also dramatically affect the Dark Ages absorption trough at z > 50. A farside array operating in total power mode from 5-40 MHz will cleanly distinguish between the standard cosmology model with adiabatic expansion and models with added cooling, possibly due to new interactions with dark matter, at >5-sigma level.

Second, spatial fluctuations in the 21-cm Dark Ages signal are governed almost entirely by well-understood linear structure formation, the same physics used to interpret *Planck* and other observations of the CMB power spectrum, allowing precise predictions of the expected signal within the standard cosmological model. Interferometric measurements of fluctuations in the 21-cm Dark Ages signal can therefore uniquely test the standard cosmological model at the onset of structure formation, without the complication of highly non-linear baryonic effects. Any departure from these well-constrained predictions will provide important new insights on the physics of structure formation, potentially into the nature of dark matter (Slatyer et al. 2013), early dark energy (Hill & Baster 2018), or any exotic physics (Clark et al. 2018). Fundamentally, such observations could also measure the ultimate number of linear modes (independent Fourier modes of the 3D density field) in the Universe and lead to exquisite cosmological constraints, including the masses of neutrinos and their hierarchy (Mao et al. 2008), the non-Gaussianity of initial density perturbation (Muñoz et al. 2015), and the imprints of primordial gravitational waves to reveal the complexity and energy scale of the inflationary paradigm (Book et al. 2012; Ansari et al. 2018).

**Redshifted Global 21-cm Signal**[1]

A promising method to measure the properties of the Dark Ages and the Cosmic Dawn in the near term is the highly redshifted Global or all-sky 21-cm spectrum illustrated in Figure 2 (Pritchard and Loeb 2012). The "spin-flip" transition of neutral hydrogen emits a photon with a rest frame wavelength of 21-cm (ν = 1420 MHz). As neutral hydrogen composed the majority of baryonic matter in the early Universe, its emission in the Global 21-cm spectrum provides crucial cosmological information from the Dark Ages through the Cosmic Dawn, as well as the EoR. Its signal is generated by a weak hyperfine transition, but its ubiquity means it can be detected even with a simple radio dipole experiment.

---

[1] See also RFI response by Burns, Bale, Bradley et al. (2021) entitled "Global 21-cm Cosmology from the Farside of the Moon" which describes a proposed pathfinder single telescope observation of the global signal in orbit of the Moon or on the lunar farside surface.

The curves in Figure 2 show broad spectral features that are common to virtually all 21-cm models of the early Universe. Its diagnostic trait, brightness temperature ($\delta T_b$) measured relative to the CMB, is driven by the evolution of the ionization fraction ($\bar{x}_{HI}$) and spin temperature $T_S$ (measure of the fraction of atoms in the two spin states) of HI relative to the radio background temperature $T_R$ (Furlanetto et al. 2006; Shaver et al. 1999): $\delta T_b \propto \bar{x}_{HI} \left(1 - \frac{T_R}{T_S}\right)$. Because of cosmic expansion, the observed frequency ν₀ corresponds to a redshift z through the relation ν/ν₀=1+z, where ν = 1420 MHz is the rest frame frequency. The greater the redshift of the hydrogen, the older it is, so $\delta T_b(\nu)$ probes the growth of structure in the early Universe as illustrated in Figure 2.

In the standard ΛCDM cosmological model, the lowest frequency (corresponding to the highest redshift and earliest time) spectral absorption feature in each curve of Figure 2 is called the "Dark Ages trough" (ν < 40 MHz). It is purely cosmological and thus relatively simple to interpret because there are no stars or galaxies at this epoch to complicate the signal. At z ≳ 30 (≲100 Myrs after the Big Bang), cosmic expansion drives a decoupling between the spin temperature and the radio background radiation temperature ($T_R$ > $T_S$) producing a broad absorption feature in the 21-cm spectrum. The standard cosmological model (dashed black curve in Figure 2) makes a precise prediction for the central frequency (≈18 MHz) and brightness temperature (≈40 mK) for this feature. Any departure from the standard model would indicate the need for additional physics such as, e.g., non-gravitational interactions between baryons and dark matter (Barkana 2018; Muñoz and Loeb 2018; Berlin et al. 2018; Muñoz et al. 2018), as shown by the color parametric curves in Figure 2. Thus, the low frequency 21-cm spectrum offers a unique and robust probe of the cosmological model.

Star formation, which probably begins around z~30, produces Ly-α photons that couple the gas temperature to the hyperfine spin-flip transition via the Wouthuysen-Field mechanism (Pritchard and Loeb 2012). As the gas adiabatically cooled faster than the CMB radiation, this results in a broad absorption feature in the 21-cm spectrum, the "Cosmic Dawn trough", starting at ≈50 MHz (Figure 2). Thus, this extremum point where the spectrum turns downward is a measure of when the first stars "turn on" and the gradient of the spectrum provides constraints on the nature of the first stars (e.g., the ratio of first generation, Pop III, to second generation, Pop II, stars; Burns et al. 2017). The massive black hole remnants of these first stars accrete gas and generate X-ray emission, which heats intergalactic neutral hydrogen and drives the 21-cm signal back towards the CMB temperature (see troughs at z~23-13, ν~60-100 MHz in Figure 2).

Recent results (Bowman et al. 2018) from the *Experiment to Detect the Global EoR Signature* (EDGES) suggest the presence of a strong feature in the 21-cm spectrum at ≈78 MHz (z≈17), within the time period expected for the Cosmic Dawn trough. Figure 2 shows examples of deviations from the standard cosmology for phenomenological models of added cooling to the primordial neutral hydrogen (Mirocha and Furlanetto 2019) that could be produced by, e.g., previously unanticipated interactions between baryons and dark matter particles. The black dashed curve assumes a standard ΛCDM cosmology model and that the earliest galaxies evolve according to smooth extrapolations of the known high-z galaxy population (Burns et al. 2017; Mirocha et al. 2016). While the difference in the observed EDGES redshift can be explained within the standard model, the drop in brightness is ≈3 times greater than allowed by adiabatic gas cooling at the frequency corresponding to the trough minimum relative to the standard model (Figure 2). The Dark Ages trough, at ν < 30 MHz and inaccessible from the ground due to ionospheric effects, is produced purely by cosmology and is thus cleaner because there are no stars to complicate the signal. This trough is a prime target for a farside low frequency array. **If the measured redshifted 21-cm signal differs from that of the standard cosmological model, new physics is required.**

There are several possible explanations for the deep Cosmic Dawn trough (grey bands) in Figure 2. First, it might be explained by an increase in the radio background above the CMB ($T_R$; Feng and Holder 2018; Fialkov and Barkana 2019), sourced by synchrotron emission in the first star-forming galaxies (Mirocha and Furlanetto 2019) or Active Galactic Nuclei (Ewall-Wice et al. 2018, 2020), or heated by neutrinos (Chianese et al., 2019), dark matter annihilation (Fraser et al. 2018), or even superconducting strings (Brandenberger et al. 2019). Second, the trough could be produced through enhanced cooling of the hydrogen ($T_S$) via, e.g., Rutherford scattering of baryons off of dark matter (Muñoz et al. 2015; Barkana 2018; Fialkov et al. 2018). However, independent constraints suggest that this source of scattering could not constitute all of the dark matter in the Universe but rather a sub-percent fraction (Muñoz and Loeb 2018; Berlin et al. 2018; Kovetz et al. 2018). **Identifying excess cooling in the early Universe could thus provide the first evidence that there is more than one kind of dark matter.** Indeed, the timing of the signal alone can constrain the properties of any warm component of dark matter (Safarzadeh et al. 2018; Schneider 2018; Lidz and Hui 2018). Third, systematic effects from the environment on Earth may also impact the purported EDGES results (Hills et al. 2018; Bradley et al. 2019; Tauscher et al. 2020; Sims et al. 2020).

In Figure 2, we show three examples of "excess cooling" models (solid color curves) that, while consistent with the EDGES trough at 78 MHz, make different predictions for the Dark Ages signal. Based upon our parametric models, these curves demonstrate the effects of both different cooling rates and time of the cooling, while adjusting astrophysical parameters in order to preserve the 78 MHz feature seen by EDGES. The 78 MHz Cosmic Dawn trough, even though suggestive of exotic physics such as dark matter interactions, describes a complicated epoch in which multifaceted astrophysics such as star formation, ionization, and black hole X-ray heating occur. The Dark Ages absorption feature, to be measured by the proposed farside array, reflects the simple state of the Universe before the formation of the first stars. Observations of the global 21-cm signal from the Moon's farside have the potential to resolve these ambiguities, to cleanly measure and constrain the origin and characteristics of any source of additional cooling, and to reveal new particle physics.

**FARSIDE Dark Ages Global Measurement**

The major challenge for Global signal observations is the presence of bright foregrounds that are $10^{4-6}$ times that of the 21-cm signal, which together with chromaticity of the antenna beam, can introduce added frequency structure into the observed foreground brightness temperature spectrum. FARSIDE mitigates these effects in several important ways. First, the array would provide high resolution maps of the sky foreground at multiple relevant frequencies. This facilitates the extraction of the 21-cm signal using a data analysis pipeline based on pattern recognition algorithms informed by training sets (Tauscher et al. 2018) that can be obtained from sky foreground observations, lab instrument measurements, beam simulations and 21-cm signal theory. Second, an orbiting calibration beacon would permit us to map the antenna beam in the far-field for the first time enabling correction for beam chromaticity effects.

The system temperature ($T_{sys}$) at these frequencies is driven by emission from the astrophysical foreground whose thermal noise is determined via the radiometer equation: $\sigma_{RMS}=T_{sys}(N\Delta\nu\cdot t)^{-1/2}$, where $\sigma_{RMS}$ is the RMS thermal noise, N is the number of antennas (used in auto-correlation mode), $\Delta\nu$ is the frequency bandwidth, and $t$ is the integration time. In our observation band, the non-thermal brightness temperature of the Galaxy increases with decreasing frequency ($T_{sys}\sim5000\text{K}[\frac{\nu}{50MHz}]^{-2.5}$). Thus, the integration time increases quickly as we observe at lower frequencies. Combining N=128 auto-correlations of the array at 15 MHz, ~1000 hours of integration time is required to achieve an RMS noise level of ≲5 mK, assuming a frequency channel width of 0.5 MHz. Such an array would thus be able to distinguish the standard cosmology from added cooling models at better than $5\sigma$ significance.

### The Dark Ages 21-cm Power Spectrum

In many ways, **spatial fluctuations in the 21-cm absorption during the cosmic Dark Ages provide the ultimate cosmological observable** (Figure 3). The simplest way to quantify these fluctuations is with the power spectrum, which characterizes the amplitude of the variations as a function of spatial scale, analogous to *Planck* measurements of the CMB. During this time, the 21-cm line traces the cosmic density field with most modes in the linear or mildly non-linear regime, allowing a straightforward interpretation of the measurement in terms of the fundamental parameters of our Universe (Lewis & Challinor 2007). **The lack of luminous astrophysical sources makes the Dark Ages signal a clean and powerful cosmological probe and also renders the 21-cm line the *only* observable signal from this era.**

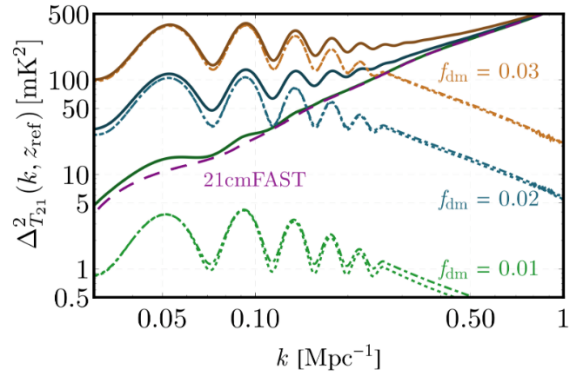

**Figure 3.** The 21-cm power spectrum can distinguish between different exotic physics scenarios during the Dark Ages. In these models, a fraction $f_{dm}$ of the dark matter is assumed to have a small charge; the oscillations in the power spectrum arise from the large-scale streaming of baryons relative to dark matter. The solid curves are the total power for each value of $f_{dm}$, after linearly adding the dash-dotted lines, showing the contributions from dark matter-baryon scattering, to the standard cosmological model (labeled "21cmFAST"). Figure from Muñoz et al. (2018).

Furthermore, the 21-cm line can be used to reconstruct a 3D volume (as compared to the 2D surface of the CMB) and is not affected by "Silk damping" on the smallest scales, which blurs out fluctuations in the CMB—meaning the number of accessible modes is enormous and is effectively only limited by the collecting area of the instrument. **Because 21-cm measurements can access an enormous number of modes in the linear regime inaccessible by other means, they enable stringent constraints on the physics of the early Universe, and in particular its early phase of inflation. It includes the running of the matter power spectrum spectral index (Mao et al. 2008), non-Gaussianity (Muñoz et al. 2015, Cooray 2006), and primordial gravitational waves (Book et al. 2012; Schmidt et al. 2014), as well as constraints on the curvature of the Universe (Ansari et al. 2018) and the neutrino masses (Mao et al. 2008). Constraints on each of these parameters would be improved by orders of magnitude, and provide unprecedented insights on the physics of inflation – its energy scale, the field(s) driving it, their interaction(s) and self-interaction (Ansari et al. 2018; Furlanetto et al. 2019). Such precision measurements would also lead to unique constraints on potentially sharp features in the inflationary potential which illuminates directly the physics at play then (Chen et al. 2016).**

The 21-cm power spectrum is also an exquisite probe of physics occurring during the Dark Ages themselves. All the physical processes that affect the Global signal described above, including the exotic scenarios proposed for explaining the EDGES signal, also affect the power spectrum. But the latter has far more information so allows more precise tests of the scenarios. Moreover, many exotic processes imprint distinct signatures in the Dark Ages power spectrum (Figure 3) (Muñoz et al. 2018). Even if the EDGES measurement is not confirmed, dark matter itself remains a mystery and **the cosmic Dark Ages offer the cleanest astrophysical probe of dark matter physics on cosmic scales.** For example, any warm dark matter will suppress the formation of small structures and hence the amplitude of the power spectrum on those scales. Constraining such scenarios with observations targeting later epochs is complicated by the slew of baryonic feedback processes that also operate on small scales. As a result, 21-cm observations targeting epochs after the Cosmic Dawn suffer a degeneracy between astrophysical processes that halt star formation in low-mass dark matter halos and cosmological processes that suppress the formation of those dark matter halos. Observations aimed at the Dark Ages avoid this important degeneracy.

**A Dark Ages Power Spectrum Pathfinder (FARSIDE) and Future Expansion**

There are many challenges to detecting the Dark Ages power spectrum. Intrinsically, it is an extremely faint signature, and sky-noise dominates any measurement (Figure 3). The signal strength predicted by the standard cosmological model will require a collecting area of ∼5 square kilometers, necessitating an array of ~100,000 dipoles to be deployed above the Earth's ionosphere—far larger than the initial array described above. Astrophysical foreground emission, including synchrotron and free-free emissions from the Galaxy and extragalactic sources, complicates the picture even further. If residual foreground emission renders some modes unusable for cosmology, even more collecting area will be required to obtain a high significance 21-cm signal detection. A pathfinder array of 128 dipole pairs, FARSIDE, is therefore an invaluable pathfinder for a larger 21-cm power spectrum instrument. It can both measure the intrinsic spectrum of the low-frequency foregrounds without the effects of the Earth's ionosphere and can be used for precision characterization of Galactic foregrounds and as a testbed for foreground removal techniques. Such measurements will be necessary for determining the array size ultimately necessary for a 21-cm Dark Ages power spectrum measurement.

The advantage of a dipole array is that is it easily expanded once the core infrastructure (power, comms, correlator, etc.) is in place. An attractive feature of the Moon is that the regolith is rich in conductive materials such as aluminum. This presents the opportunity to expand the lunar array to 100,000 dipoles by processing lunar regolith into raw materials, and then fabricating, via vacuum vapor deposition, the RF antennas, electric power generation system, electrical energy storage, and other components needed to create this unprecedented cosmology observatory[2]. The key enabling technology for this observatory is a Molten Regolith Electrolysis (MRE) process, a high-temperature electrolytic process in which the naturally high-oxide lunar regolith is liquefied and by the action of electric current is dissociated to electroextract liquid metal as a product and oxygen as a by-product (Sadoway et al 2019). MRE needs only power, oxides, and gravity to extract oxygen and metals from regolith. In addition to manufacturing the dipoles, this technology can also extract the needed materials to manufacture solar arrays, batteries, and the electrical wires to connect the dipoles. The technology can produce the observatory within 2-3 years and remain active to maintain, refurbish, and upgrade the observatory for decades.

**RELEVANCE TO NASA ASTROPHYSICS & DOE HIGH ENERGY PHYSICS PRIORITIES**

A Dark Ages 21-cm array on the lunar farside addresses NASA's objectives in astrophysics to "Discover how the universe works, explore how it began and evolved" and to "Probe the origin and destiny of our universe, including the nature of black holes, dark energy, dark matter and gravity." It will execute a mission recommended in the NASA Astrophysics Roadmap to observe "the Universe's hydrogen clouds using 21-cm wavelengths via observations from the farside of the Moon", in part to "Resolve the structure present during the Dark Ages and reionization epochs." The recent Artemis III Science Definition Team (SDT) Report advocated for observations from "the farside of the Moon and the 21-cm electromagnetic radiation spectral line to study the Dark Ages of the universe."

The NRC Astro2010 Decadal Survey, New Worlds, New Horizons in Astronomy and Astrophysics, identified the Dark Ages and Cosmic Dawn as one of cosmology's great frontiers. This array will begin to tackle a key question posed by the survey: "What is the nature of dark matter?".

Redshifted 21-cm observations of the Dark Ages are consistent with the mission of DOE HEP "to understand how our universe works at its most fundamental level". These observations support three of the Science Drivers of Particle Physics including "identify the new physics of dark matter," "understand cosmic acceleration: dark energy and inflation," and "explore the unknown: new particles, interactions, and physical principles."

---

[2] Recently, NASA selected this concept for a 2021 Phase I NIAC study: https://www.nasa.gov/directorates/spacetech/niac/2021_Phase_I/FarView/

## 2.  ENGINEERING DESIGN FOR A FARSIDE ARRAY

NASA funded the first engineering design study of an interferometric array for deployment by a large CLPS[3] lander on the lunar farside (Burns et al. 2019, 2021). FARSIDE[4] is a potential example of NASA Astrophysics' new mission class called Probes which are PI-led missions with a funding cap of ≈$1 billion. The FARSIDE architecture and operations concept were developed via a series of studies at JPL. These studies assessed the technical feasibility of the instrument and the mission, and the cost-realism for a Probe-class mission. The current FARSIDE design architecture consists of 128 non-cospatial orthogonal pairs of antenna/receiver nodes distributed over a 10-km✕10-km area in a four-arm spiral configuration shown in the figure on the cover page. The FARSIDE mass is ∼1800 kg delivered to the lunar surface by a large commercial lander[5]. Four two-wheeled tethered rovers, such as JPL's Axel (Nesnas et al. 2012), would be teleoperated to deploy the nodes on the four-arm spiral, which remain connected to the lander by their individual tethers, providing communication, data relay, and power during deployment as well as for operations.

The array will operate over a frequency range of 0.1-40 MHz, with ∼5-40 MHz used for Dark Ages cosmology. At its lowest frequency, FARSIDE would operate two orders of magnitude below the Earth's ionospheric cutoff. In the low-frequency band, the array would image the sky visible from the lunar surface once per minute to search for signatures of space weather (e.g., Type II and III solar radio bursts) and planetary magnetospheres, from solar and exoplanetary systems (Burns et al. 2021).

The array sensitivity is determined by the number and type of antennas, and the resolution is governed by the maximum baseline. For FARSIDE, we use 10-100-m tip-to-tip thin-wire antennas. It was determined that 128x2 dipole antennas will provide the needed sensitivity to span the science cases described in Burns et al. (2019). With an array area of 10-km✕10-km, FARSIDE achieves a resolution of 10 arcmin at 15 MHz.

Sensitivity specifications for FARSIDE are summarized in Table 1. The array sensitivity depends upon the frequency bandwidth, the integration time, the system temperature, and the effective area of the array. The effective area is determined by the dipole length and the antenna impedance. The system temperature depends upon the sky and regolith temperatures, and the front-end amplifier. See Burns et al. (2019) for complete details of the modeling and simulations.

**Table 1.** FARSIDE Sensitivity at 15 MHz

| Quantity | Value |
|---|---|
| Frequency Coverage | 0.1 – 40 MHz (1400 ✕ 28.5 kHz channels) |
| System Temperature ($T_{sys}$) | $2.7 \times 10^4$ K |
| Effective Collecting Area ($A_{eff}$) | 2240 m$^2$ |
| System Equivalent Flux Density ($2k_B T_{sys}/A_{eff}$) | $2.8 \times 10^4$ Jy |
| 1$\sigma$ Sensitivity for 1 hour, $\Delta\nu = \nu/2$ | 120 mJy |

The FARSIDE front end uses the electrically short antennas to achieve sky background noise-limited observations. On the lunar farside, the lunar highlands are thick, have low conductivity, and vary slowly with depth, thus removing the need for a ground plane. The antennas will be placed directly on the lunar surface. The net impact of this multipath effect is to introduce a direction-dependent component to the array synthesized beam. FARSIDE will use an orbital beacon for calibration and to map both the antenna beam pattern and the array synthesized beam to fully capture this effect. The array leverages existing designs from front-end amplifiers, fiber optics, and the correlator system based upon ground-based observatories (e.g., OVRO-LWA, Anderson et al. 2018).

The concept design calls for operations during the lunar night and day using proven low-temperature electronics and batteries. Recent tests indicate that resistors and capacitors, forming the basis of the electronic components of the receiver nodes, are capable down to temperatures <-180°C (Ashtijou et al. 2020).

---

[3] Commercial Lunar Payload Services, https://www.nasa.gov/content/commercial-lunar-payload-services-overview.
[4] FARSIDE is an acronym for **F**arside **A**rray for **R**adio **S**cience **I**nvestigations of the **D**ark ages and **E**xoplanets.
[5] Example is Blue Origin's Blue Moon lander. See https://www.blueorigin.com/blue-moon/

With four Axel rovers, the antenna nodes will be deployed along four-spiral arms, and hence the full array, in <1 lunar day. The nodes are embedded within each of the four tethers. As such, the deployment of the 128 node pairs requires no electromechanical mating at any point during the mission. The tether, with its embedded nodes, is hosted on the rover and is paid out by a spool on the rover as it drives along the spiral. A landing site will be selected that is relatively benign for the rover deployment operation.

The lander includes solar panels and batteries to provide power for the receivers and the signal processor using an FX correlator (e.g., Kocz et al. 2014), as well as communication of the data back to Earth. The solar panels charge the rovers' batteries via their respective tethers during deployment and power the nodes during operations. After cross-correlation, visibility data will be conveyed to Earth for further analysis. This is estimated to be about 130 GB of data per 24 hour period.

## 3. RESPONSES TO RFI QUESTIONS

- **Science Topics**

a. What are key science topics aligned with HEP and/or APD science drivers?

A 21-cm Cosmology radio telescope on the lunar farside, aiming at the global (spatially-averaged) signal, opens a new frontier to the early Universe in the Dark Ages. It would allow unprecedented tests of the standard $\Lambda$CDM cosmology model in an uncharted epoch, and the potential revelation of new physics involving dark matter, early Dark Energy, and other exotic physics. A large-scale 21-cm Cosmology radio array on the surface of lunar farside, aiming at the three-dimensional spatial fluctuation signal, would further provide the most exquisite probes in fundamental physics and cosmology in the observable Universe, shedding light on the energy scale and complexity of inflation, neutrino species, among others. These unique science opportunities match well with both NASA and DOE science drivers to "understand how the Universe works at a fundamental level". We propose a pathfinder array on the lunar farside, FARSIDE, that would be capable of accurately characterizing the 21-cm global signatures in the Dark Ages, testing the standard cosmology model in the highest redshifts after recombination and quantifying any deviation due to new physics. FARSIDE would further enable feasibility studies of the 21-cm fluctuation measurements on limited spatial scales, allowing characterization of the lunar radio environment and astrophysical foreground emissions in these low-frequency range that is otherwise inaccessible from earth. Development of mitigation strategies and analysis technique would be invaluable to further the science goals.

b. Are precursor scientific measurements or demonstrations needed?

High fidelity Global signal measurements of the Cosmic Dawn epoch are needed to demonstrate measurement feasibility in preparation for the more challenging Dark Ages observations and to resolve current uncertainties regarding the magnitude of the Cosmic Dawn absorption trough. A single telescope experiment between 40-110 MHz on a NASA CLPS[6] farside lander would be ideal as it eliminates systematics that plague Earth-based experiments, including radio frequency interference, ionospheric absorption and refraction, and ground/atmosphere environmental variations (see RFI response by Burns, Bale, Bradley et al.).

c. What key obstacles, impediments, or bottlenecks are there to advancing the scientific research?

More theoretical work is needed to robustly forecast and simulate the wide range of 21-cm global and fluctuation signals based on alternatives or extensions to the minimal cosmology model, including warm or fuzzy dark matter, early Dark Energy, and cosmic strings. What sensitivity, and spectral and spatial resolution are required to separate these various exotic physics models? These predictions will then feed into the design of the farside array. Foreground contamination to the 21-cm signal, including human-generated radio frequency interference (RFI) and astrophysical radio synchrotron and free-free emission associated with the Galaxy and extragalactic sources, overwhelm the 21-cm signals by 4-6 orders of magnitudes, depending on the frequency. Proper mitigation of foregrounds and quantification of the residual is key to the success of the program, which relies on careful design and characterization of the array properties. In addition, analogous to the successful field of CMB, developing innovative analysis techniques using e.g., Bayesian inference, end-to-end numerical simulations, parameter extraction, and science interpretation would be critical to maximize the potential science return.

---

[6] CLPS = Commercial Lunar Payload Services (https://www.nasa.gov/content/commercial-lunar-payload-services)

- **Technology Capabilities –**

d.  What are existing or near-term technology capabilities available to advance science goals?

One of the key technology capabilities is the deployment of long tethers with embedded antenna nodes via remote operations and the post-deployment operation of antenna node pairs on the lunar surface. JPL Axel rovers have demonstrated the deployment of hundreds of meters of tether across a range of relevant terrains via remote operation. But, such deployment did not include embedded antenna nodes. The development and demonstration of an appropriate-scaled prototype that can deploy a tether with antenna nodes, first at an Earth-analog site and, subsequently, on the lunar surface, would retire risks of the full-scaled system to cover an area of 10-km×10-km. The specific capabilities to be developed and matured for FARSIDE include:

1. **Tether/node design:** prototype, demonstrate and test the viability of a tether-embedded node design using low-temperature electronics to survive lunar nights.
2. **Rover design:** adapt the design of the Axel rover to deploy long tethers with multiple embedded nodes.
3. **Remote antenna deployment:** build a rover prototype capable of the above deployment and demonstrate in relevant environment on Earth using remote operations.
4. **Tether assessment and handling:** develop tools to assess proper tether deployment to identify early indicators of long-length deployment challenges and mitigation strategies.
5. **Antenna/tether memory:** characterize challenges associated with antenna memory and develop mitigation strategies.

An Earth prototype followed by a lunar demonstration on a CLPS lander would include a single Axel rover that egresses from the lander, deploys a ~100 m antenna with 4-6 nodes using a multi-reel spool. The key risks that such a demonstration would retire include: multi-reel spools with stem nodes, deployment of antenna/nodes on lunar surface without entanglement, night operation of nodes using low-temperature electronics, scaled demonstration of antenna function, remote antenna deployment and rover operation via tether (power/comm).

e.  Are precursor technology developments or demonstrations needed?

A precursor demonstrator interferometer on the lunar surface is needed to reduce technology risks. For example, a six-element radio array could operate over a frequency bandwidth of 0.1-20 MHz and a maximum baseline of 200-m. Use of a small rover will demonstrate the deployment of receiver/antenna nodes on the lunar surface. The rover will use a tether connected to the lander for power, communications, and data transmission. Thin wire, electrically-short, 10-m antennas would be embedded within the tether, following the design concept developed for FARSIDE. Investigation of the electrical properties of the antennas placed directly on the dielectric regolith will characterize the impact of the reflection of radio waves from the subsurface, informing requirements for the characterization of this effect via an orbiting calibration beacon for FARSIDE. The electronics used in the antenna nodes will have heritage from NASA's heliophysics mission SunRISE[7], a six cubesat interferometer in Earth orbit to measure compact radio structure in solar Coronal Mass Ejections. In addition, this prototype array will test the thermal designs and baseline concept of operations. The technology objective is to validate the instrument concept by demonstrating background noise-limited performance. Such an array will have sufficient sensitivity to produce dynamic spectra for transient emission from the Earth (if emplaced on the nearside), the Sun, and Jupiter.

---

[7] https://www.colorado.edu/ness/projects/sun-radio-interferometer-space-experiment-sunrise

f.  What are the key obstacles, impediments, or bottlenecks?

While the technology for FARSIDE is relatively mature, there is some development required to ensure the technology can be scaled up to a future array of 100,000 antennas. Such an array would occupy a similar footprint to FARSIDE, but with a much denser array of smaller antennas (10 m) over a narrower frequency range with singular focus on Dark Ages science. The cost and mass would scale at a rate much less than linear - for example, the total tether size is ~80x longer than for FARSIDE for a 800x larger number of antennas. This assumes a compact node design with low temperature electronics and a scalable platform for digital signal processing.

**Low-temperature Electronics** - The lunar farside is an optimum environment for isolation from anthropogenic RFI without ionospheric corruption that prevents similar Earth-based experiments. To maximize the utility of this site, night-time observations are required, preventing contamination from the Sun which can be time variable and present frequency structure that is a contaminant to the cosmological signal. This requires the use of electronics for the antenna that can survive the lunar night (~-170℃). Development of low-temperature electronics within a small form-factor node design is a key step for both FARSIDE and a next-gen power-spectrum array.

**Digital Signal Processing** – The FARSIDE array uses a complex correlator to interfere the signals of each antenna to produce the visibility data that are sent to Earth. For FARSIDE, the correlator makes use of the JPL-developed, radiation-tolerant, FPGA-based, Miniaturized Sphinx onboard computer. A next-generation array would require a much more powerful complex correlator that may dominate the cost of the mission without a new platform to lower cost and power requirements. A complex correlator scales with the square of the number of antennas (N), such that a naïve scaling from FARSIDE would imply a factor of 500,000x increase in the compute requirements for the correlator. This can be mitigated somewhat by use of a specialized Modular Optimal Frequency Fourier (MOFF) correlator (Morales 2008) where the scaling is $N*logN$, resulting in an increase of 5,000x relative to FARSIDE. This may be feasible through use of a custom application-specific integrated circuit (ASIC) instead of an FPGA, designed for low cost and power consumption.

- **Collaborations**

g.  What cooperation or partnerships could further the scientific and technology advances?
h.  What is the mix of institutions or collaboration models?

The science opportunities and data analysis challenges offered by 21-cm lunar measurements closely resemble (and would surpass) the CMB field; there has already been successful and on-going collaboration between DOE and NASA labs, such as the Planck US collaboration and the CMB-S4 collaboration. We envision a similar partnership for lunar 21-cm exploration, where NASA would provide the unique space, landing, deployment and rover technology, and DOE the large-scale experimental capability for procurement, operation, high-performance computing, and science collaboration. Both NASA and DOE would engage with their university partners for scientific collaboration, as well as the science community as a whole, while closely participating in the planning, building, and operation of the lunar experiment(s) on the surface or in orbit.

i. What resources, capabilities and infrastructure at DOE and NASA would be beneficial?

JPL has already invested in several aspects of lunar exploration, as the lead center for the FARSIDE Probe Study and sponsor of several NASA Innovative Advanced Concepts (NIAC) lunar concepts. It has a track record of successful collaboration with DOE laboratories in the context of the past Planck mission, and the on-going Dark Energy Survey (DES) and CMB-S4 project. JPL also has unique leadership in solar system robotic surface exploration, e.g., the Mars rovers, and in large data analysis and numerical simulation challenges (Planck, DES, Euclid, Roman). These are key aspects necessary to facilitate successful lunar 21-cm experiment(s) and could accelerate the research progress.